\documentclass[pre,twocolumn,showpacs,amsmath,showkeys]{revtex4}

\usepackage{graphicx}
\usepackage{bm}

\begin{document}

\title{Adiabatic invariance with first integrals of motion}

\author{Artur B. Adib}
\email{artur.adib@brown.edu}
\affiliation{
  Department of Physics and Astronomy, Dartmouth College,
  Hanover, NH 03755, USA
}
\altaffiliation[Present address:]{
  Box 1843, Brown University,
  Providence, RI 02912, USA.
}

\date{\today}

\begin{abstract}
The construction of a microthermodynamic formalism for isolated systems based on the concept of
adiabatic invariance is an old but seldom appreciated effort in the literature, dating back at least
to P. Hertz [Ann. Phys. (Leipzig) {\bf 33}, 225 (1910)]. An apparently independent extension of such formalism
for systems bearing additional first integrals of motion was recently proposed by Hans H. Rugh
[Phys. Rev. E {\bf 64}, 055101 (2001)], establishing the concept of adiabatic invariance even
in such singular cases. After some remarks in connection with the formalism pioneered by Hertz, it will
be suggested that such an extension can incidentally explain the success of a
dynamical method for computing the entropy of classical interacting fluids, at least in some potential
applications where the presence of additional first integrals cannot be ignored.
\end{abstract}

\keywords{Ergodic adiabatic invariance; Adiabatic switching method; Molecular dynamics ensemble}

\pacs{05.20.Gg, 05.45.-a, 05.70.-a, 02.40.Vh}

\maketitle


Consider an isolated collection of particles with $f$ degrees of freedom evolving in time according
to the usual laws of classical particle mechanics  (e.g., a gas
of $N=f/3$ particles inside a closed cylinder). The dynamics of this system can be uniquely described
by a trajectory in the $2f$-dimensional phase space of the canonically conjugate variables
${\mathbf q}=(q_1,...,q_f)$ and ${\mathbf p}=(p_1,...,p_f)$. Let $H({\mathbf q}, {\mathbf p}; \lambda)$ be
the Hamiltonian of the system depending (for simplicity) on a single external parameter
$\lambda$ (e.g., the volume set by the position of the piston in the cylinder). Assume
further that no other first integrals exist (see below, however). Then, for constant $\lambda$,
the trajectory of the system will be confined to and will (almost) fill out the whole surface of
constant energy $H({\mathbf q}, {\mathbf p}; \lambda)=E_\lambda$, provided ergodicity is assumed.
Consider now a process in which $\lambda=\lambda(t)$ varies very slowly in comparison to
a typically small ``observation time'' $\tau$, which in turn is a sufficiently large quantity with regard to
microscopic processes such that time averages of phase space functions taken over it approximate the 
corresponding microcanonical averages at constant $\lambda$ (this is an idealized instance of an
otherwise precisely observed process, e.g. a reversible compression/expansion of the gas in the
cylinder above). Such
 processes are usually referred to as {\em adiabatic} (though an
additional {\em quasistatic} qualifier would be certainly very welcome), and it
was shown by P. Hertz in his papers titled ``On the Mechanical Foundations of Thermodynamics'' 
\cite{hertz10a,hertz10b} that the ($\tau$-averaged) phase space space volume $\Omega(E,\lambda)$ enclosed 
by the surfaces of constant energy $H({\mathbf q}, {\mathbf p}; \lambda)=E_\lambda$
is a conserved quantity during processes of this kind (justifying then the label {\em adiabatic invariant}),
much like the macroscopic entropy of Clausius under the same conditions.
Thermodynamics could thus be constructed from mechanical arguments, leading directly to a microscopic form
of entropy, $S_\Omega=\ln \Omega$, which should be contrasted to the usual form inspired by Boltzmann's
ideas involving the phase space area $\omega=\partial \Omega/\partial E$, $S_\omega=\ln \omega$, the
difference between these two being particularly relevant for finite systems \cite{berd-jamm88,umirzakov99,adib02c}.
These results of Hertz were greatly appreciated by Einstein (see \cite{einstein11hertz} in particular)
and are often taken as the starting point of statistical mechanics in
the German literature \cite{munster69,becker67}. However, these simple and elegant observations seem to have
escaped the attention of modern \cite{huang87,pathria96,reichl98} and even
classic \cite{landau80-1} treatments (some rare exceptions are Refs. \cite{kubo65,berd-book97}).
The underlying derivation is simple and accessible and thus will not be reproduced here (this so-called
{\em ergodic adiabatic invariance} problem, however, is far from being exhausted by Hertz's original papers,
being constantly addressed at different levels in the literature, see e.g.
\cite{kasuga61,ott79,lochak88,jarzynski93}). It suffices to mention here that, in connecting Rugh's derivation
with Hertz's results, an infinitesimal change of the parameter $\lambda$ under the presence of
{\em parameter-independent} first integrals $F_i({\mathbf q}, {\mathbf p})=I_i$ maintains
straightforwardly the adiabatic conservation of
the phase space volume $\Omega(E,I,\lambda)$, with $I\equiv\{I_i\}$, and thence of the entropy $S_\Omega$,
this observation following immediately from the fact that $(\partial \Omega/\partial I_i)\,dI_i=0$ under a
change $d\lambda$ (compare, for example, Eq. (I 236) of Ref. \cite{munster69} or Eq. (34.4) in
Ref. \cite{becker67}). Another important result presented in Ref. \cite{rugh01} was the possibility
of computing the ``bulk'' temperature by means of a microcanonical (and therefore temporal) average
at constant $E,I$ and $\lambda$, namely $T_{\Omega}=\langle {\mathbf Y} \cdot \nabla H({\mathbf Y})|E,I,\lambda \rangle$,
with ${\mathbf Y}$ parallel to the surfaces $F_i({\mathbf q}, {\mathbf p})=I_i$ and satisfying
$\nabla \cdot {\mathbf Y}=1$. I note in passing that a similar form of this
generalized equipartition theorem was obtained by M{\"u}nster \cite{munster69} for the
case of arbitrary cyclic coordinates present in the Hamiltonian (which when cast in terms of appropriate
variables, e.g. center-of-mass for the conservation of linear and angular momentum or action-angle
for more general integrals, translates essentially into the above requirement of parallelism).

Finally, I would like to bring attention to the fact that the extension provided by Rugh of Hertz's original ideas
can incidentally explain the success of the so-called ``adiabatic switching method'' \cite{watanabe90}
for computing entropies of classical interacting fluids (which is one of the very few works to make explicit
use of Hertz's results), at least for the applications discussed below. This method
takes full advantage of the adiabatic invariance of $S_{\Omega}$ by initially considering a ``reference'' system
with Hamiltonian $H_0$ whose entropy is known explicitly (e.g., an ideal gas) and slowly turning on the
interactions such that the final Hamiltonian equals the desired one, $H_1$. If
this process is sufficiently slow, it can be considered as adiabatic (in the sense defined above),
the phase space volume $\Omega(E,\lambda)$ (and thus the entropy $S_\Omega$) being a conserved quantity throughout
the whole switching process \footnote{Although the use of an ideal system is incompatible with ergodicity
and thus, in principle, should prevent the method from working, surprisingly neither ergodic
time-scales nor ergodicity itself seems to be essential, suggesting that the latter is not a necessary but
rather a sufficient condition for its success \cite{watanabe90}.}. Therefore, a thermodynamic quantity which is not a phase space function and hence not
immediately obtained by means of the usual microcanonical/temporal average \cite{haile92} can be in
principle easily computed. However, this method is usually applied in the ``molecular-dynamics ensemble''
in which the total linear momentum ${\mathbf P}$ is exactly conserved due to the use of periodic boundary conditions
(see, e.g., \cite[sec. 2.10]{haile92} and references therein). Such additional first integrals of motion should be 
carefully considered when using the Hertz invariant, since the manifold accessible to the system clearly does not 
coincide with the whole constant-energy surface. This fact seems to have passed unnoticed by the authors of Ref.
\cite{watanabe90}, although the above observations by Rugh really give a strong theoretical support to their
method. Indeed, it is easy to see that all three components of ${\mathbf P}$ are parameter-independent in this 
case: the switching on of the internal interactions (which depend on the relative position of the particles only) 
clearly does not break the translational invariance of the Lagrangian. This same idea can be equally well 
applied to a more correct formulation of this ensemble that only recently was realized \cite{ray99,wood00}, in 
which three additional first integrals (the components of the center of mass ${\mathbf R}$) are considered. 
It is worth emphasizing that this simple result was only possible because of the parameter-independence
of the integrals of motion. The opposite and more general case, however, is still an open issue, as pointed
out by Rugh.

It is my pleasure to thank V. L. Berdichevsky, C. Jarzynski and W. P. Reinhardt for
the extremely fruitful correspondences. Financial support and computational resources were provided
by Dartmouth College and its Field Theory/Cosmology group, respectively.

\end{document}